\documentclass{Interspeech2024}

\interspeechcameraready 
\usepackage{float}

\title{DreamVoice: Text-Guided Voice Conversion
\thanks{$^{\dagger}$Indicates equal contribution.}
\thanks{$^{\star}$This work was supported in part by ONR N00014-23-1-2050 and N00014-23- 1-2086.}
}

\name[affiliation={1,\dagger}]{Jiarui}{Hai}
\name[affiliation={1, \dagger}]{Karan}{Thakkar}
\name[affiliation={1}]{Helin}{Wang}
\name[affiliation={2, 3}]{Zengyi}{Qin}
\name[affiliation={1, \star}]{Mounya}{Elhilali}

\address{
  $^1$Electrical and Computer
Engineering, Johns Hopkins University, Baltimore, MD, USA\\
  $^2$Massachusetts Institute of Technology, Cambridge, MA, USA \\
  $^3$MyShell.ai, USA}
\email{
jhai2@jhu.edu, kthakka2@jhu.edu, hwang258@jhu.edu,
qinzy@mit.edu,
mounya@jhu.edu
}


\begin{document}

\maketitle

\begin{abstract}
Generative voice technologies are rapidly evolving, offering opportunities for more personalized and inclusive experiences. Traditional one-shot voice conversion (VC) requires a target recording during inference, limiting ease of usage in generating desired voice timbres. Text-guided generation offers an intuitive solution to convert voices to desired ``\textit{DreamVoices}" according to the users' needs. Our paper presents two major contributions to VC technology: (1) DreamVoiceDB, a robust dataset of voice timbre annotations for 900 speakers from VCTK and LibriTTS. (2) Two text-guided VC methods: DreamVC, an end-to-end diffusion-based text-guided VC model; and DreamVG, a versatile text-to-voice generation plugin that can be combined with any one-shot VC models. The experimental results demonstrate that our proposed methods trained on the DreamVoiceDB dataset generate voice timbres accurately aligned with the text prompt and achieve high-quality VC.

\end{abstract}

\section{Introduction}

%



The emergence of augmented reality devices and accessible virtual environments marks a significant technological shift \cite{hamad2022virtual, hupont2023next}. This transformation underscores the need to develop tools that enhance user experience in these virtual spaces, ensuring safety and engagement. Therefore highlighting the need for more intuitive interaction methods with such technologies. Imagine a world where a user (source) can effortlessly modify their voice to suit their digital persona (target), from specifying \textit{``a young male voice with a dark tone and smooth texture"} to customizing auditory experiences like making \textit{``player A's voice less harsh"}. This technology holds particular significance for individuals with gender dysphoria or speech impairments, offering them avenues for expression that were previously inaccessible.

An essential aspect of VC is providing the model with a robust and accessible representation of the target voice during inference or training. Traditionally, one-shot VC models rely on the availability of target recording to extract pre-trained speaker embeddings \cite{sisman2020overview, qian2019autovc, qian2020f0, nercessian2020improved} during inference. However, the accessibility of the target recording or embeddings is not always feasible for all applications. Recently, there has been a shift towards using text-guided control or conditioning for generative audio tasks like expressive Text-to-Speech \cite{shimizu2023prompttts++, zhang2023promptspeaker}, Text-to-Audio \cite{huang2023make} and Style Transfer \cite{yao2023promptvc}. Ultimately, the shift towards text-guided control, while offering scalability and flexibility, hinges critically on the quality of text annotations. 

Previous research studies \cite{yao2023promptvc, shimizu2023prompttts++, zhang2023promptspeaker} have attempted to annotate voice timbre, also recognized as tone or color of voice, using text-based methods. However, these datasets often face limitations such as small scale, synthetic markings, restricted access, or opaque collection strategies. PromptTTS++ \cite{shimizu2023prompttts++} employs a keyword-based marking strategy on a subset of speakers in the LibriTTS dataset \cite{koizumi2023libritts}. However, the annotation details are not clearly stated and the annotated data has not been released to the best of our knowledge.
Promptspeaker \cite{zhang2023promptspeaker} uses an semi-synthetic Mandarin dataset, of which only the internal subset of 74 speakers contains detailed timbre annotations and the remaining data merely contains gender or age information, based on the actual ground truth labels, rather than the perceived timbre.
PromptVC \cite{yao2023promptvc} uses an internal Mandarin dataset that only contains six speakers and lacks detailed documentation on the control of the text annotations of style and timbre. 
Hence, creating a comprehensive, meticulously detailed, and open-source dataset of text annotations will play a pivotal role in advancing text-based voice control and conditioning.




This study explores the task of text-guided voice generation and conversion, unlike \cite{yao2023promptvc, shimizu2023prompttts++} that use text-guidance for speech content control and generation. Our main contributions \footnote{\href {}{Demos and source code: https://research.myshell.ai/dreamvoice}} are summarized as follows:

\begin{enumerate}
    \item We release DreamVoiceDB, an extensive, open-source voice timbre dataset of 900 speakers sampled from LibriTTS-R \cite{koizumi2023libritts} and VCTK \cite{Yamagishi2019VCTK} dataset, annotated by speech and language experts for high-quality research applications. 
    \item We propose two text-guided voice generation and conversion models: DreamVC, an innovative text-guided voice timbre conversion model using Diffusion Probabilistic Models (DPM) \cite{ho2020denoising} and Classifier-Free Guidance (CFG) \cite{ho2022classifier} for effective condition controllability; DreamVG, a light and plug-and-play text-to-voice generation plugin model compatible with one-shot VC models. DreamVG also uses DPM with CFG to generate speaker embeddings.
    \item We experimentally demonstrate that the proposed dataset and models can achieve high-quality voice conversion with timbres that are precisely aligned with the text description.
\end{enumerate}







\section{DreamVoiceDB: Voice Timbre Dataset}
\label{sec:dreamvoicedb}

\begin{figure}[t]
  \centering
  \includegraphics[width=7cm]{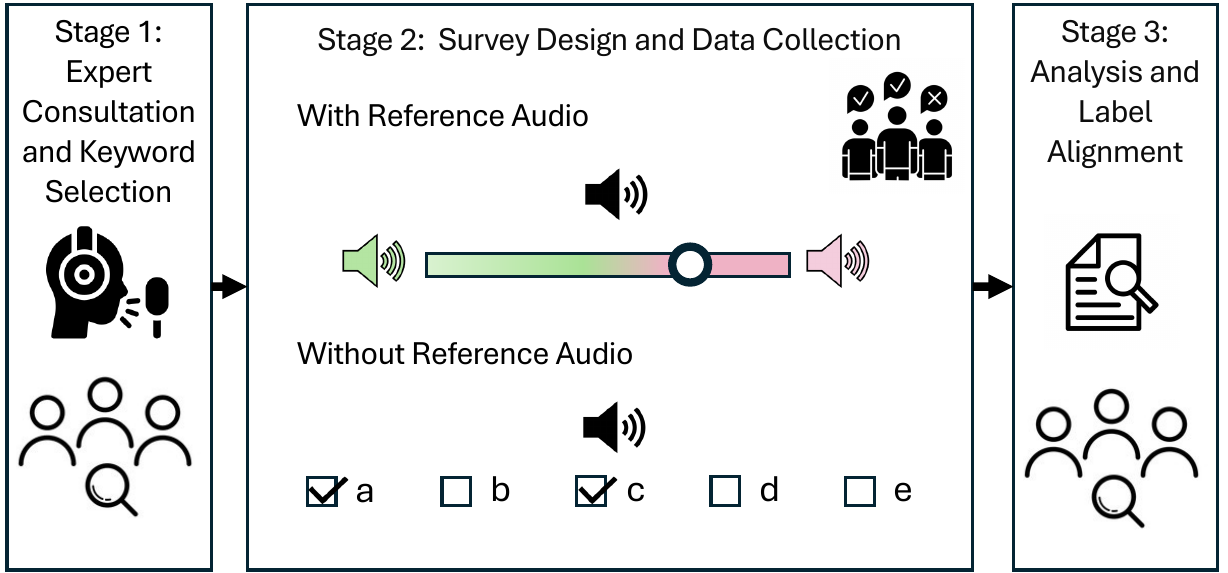}
  \caption{Schematic diagram of DreamVoiceDB survey method.}
 \vspace{-6mm}
  \label{fig:surevey_method}
\end{figure}

Voice timbre emerges from a combination of factors, including age, gender, physical properties of the vocal tract and vocal cord, and perceptual characteristics.
To accurately capture the rich characteristics of timbre we followed a comprehensive three-stage process as summarized in Figure \ref{fig:surevey_method}.           

In the first stage, 900 speakers were sampled from existing multi-speaker datasets LibriTTS-R \cite{koizumi2023libritts} and VCTK \cite{Yamagishi2019VCTK}. Following this, an expert voice actor guided us through the selection of keywords that best represent voice timbre. A total of 10 keywords were split into two categories based on their level of subjectivity. The first category focuses on basic, more objective timbre aspects like age, gender, brightness, and roughness, while the second encompasses subjective characteristics such as perceived strength, warmth, and authority. In addition to these keywords, we extended our inquiry to include the perceived voice's suitability of voice-related professions such as Storytelling and Client Interaction, linking qualities that make each voice distinct and memorable.



The second stage involved combining the expert's knowledge of the keywords deeply into our survey methodology. Questions were designed with reference audio examples to facilitate objective category annotations based on relative comparison. Assessment of subtler attributes like brightness and roughness was conducted using a Likert scale defined based on expert knowledge. The second category responses were measured using a binary scale for all the keywords in that category in compliance with their subjective nature. Following the survey design and release, a test run was conducted to recruit the best annotators. A total of 8 expert annotators, comprising 4 females and males, were selected for the study. Annotator's expertise spanned speech-related fields such as speech and language pathology, speech and accent coaching, singing coaching, and transcription work. All 900 speakers were annotated once by each expert annotator. 

Lastly, a comprehensive procedure was employed after data collection to align and investigate the identified keywords, prioritizing those based on their respective agreement scores. Keywords that garnered unanimous consensus among annotators were seamlessly integrated into the dataset. Conversely, keywords exhibiting moderate agreement levels were subjected to rigorous reassessment based on their agreement distribution and combination of manual self-reported scores. This process significantly augmented the dataset’s precision and authenticity, keeping in mind the richness and diversity of the dataset. Following keyword annotations, we used OpenAI's GPT4 \cite{achiam2023gpt} API to generate approximately 50 natural language descriptors for each speaker depending on the combinations of the keywords. The prompts generated included all leave-one-out and leave-many-out combinations to cover a wide range of practical inputs. Further details about the keyword distribution and analysis code are available \footnote{\label{note1} \href {}{Dataset and data analysis: https://research.myshell.ai/dreamvoice}}.


\section{Method}

\begin{figure*}[t]
  \centering
  \includegraphics[width=\linewidth]{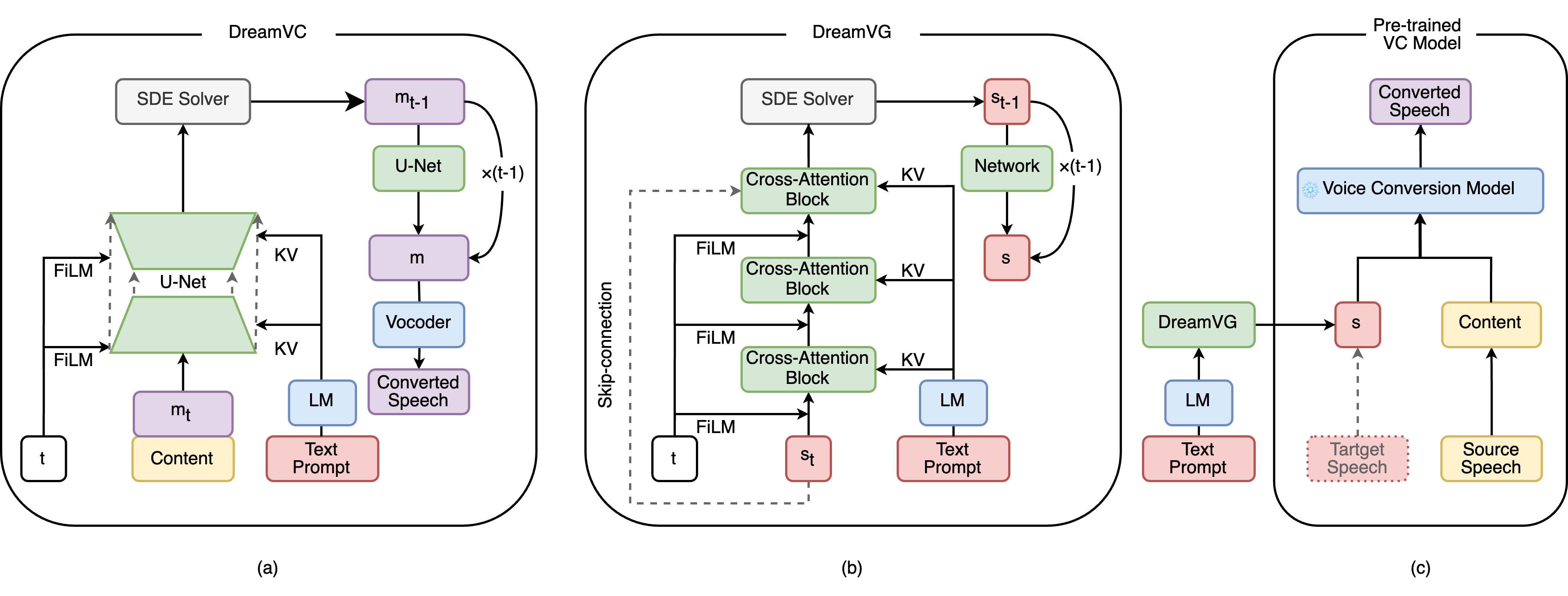}
  \caption{Overview of the (a) DreamVC, (b) DreamVG, and (c) Plugin Strategy. Modules in blue are pre-trained models and remain frozen during training, while modules in yellow are trained. Green blocks represent the source speaker information while red blocks represent the target speaker information. Purple blocks correspond to the converted speech. Dashed lines represent skip connections. LM represents the Language Model. $KV$ represents Cross-Attention \cite{vaswani2017attention} and FiLM represents Feature-wise Linear Modulation layers \cite{perez2018film} used for fusing Text Prompt and diffusion step $t$ respectively. SDE solver is the stochastic differential equations for the diffusion sampling. Text Prompt is the text description about the desired target voice. $t$ is the diffusion step. $content$ is the content embedding of the source speaker. $s$ is the speaker embedding of the target voice. $m$ is the mel-spectrogram. $m_t$ and $s_t$ represent the noisy versions of the mel-spectrogram and the speaker embedding at the diffusion step $t$. }
  \label{fig:model}
\end{figure*}


\subsection{General Voice Conversion Pipeline} 
Voice conversion models work by taking the content of the source speech, mixing it with the target speaker's timbre, and then generating the converted voice. Recent VC methods often use latent features extracted from large pre-trained Speech Language Models (SLMs), which contain limited speaker information but rich content information \cite{qian2022contentvec}, as the content embedding for the source speaker. In the case of one-shot VC, a pre-trained speaker verification model, trained on datasets with a large number of speakers, is utilized to extract the speaker embedding of the target speaker. The content embedding and the speaker embedding are then used to condition a VC model to synthesize speech with the content of the source speaker and the voice timbre of the target speaker. Researchers have deployed discriminative \cite{qian2019autovc} and generative models for this task including GANs \cite{hsu2017voice, kameoka2018stargan} and Diffusion models \cite{popov2021diffusion, li2023freevc}. While GANs \cite{saxena2021generative} are fragile in convergence and relatively hard to train, Diffusion \cite{yang2023diffusion} models provide a more stable alternative for training generative VC models. In addition, recent diffusion have shown promising performance in both diversity and quality on various text-guided generation tasks. 

\subsection{Diffusion Models and Classifier-free Guidance}

Diffusion Probabilistic Models (DPMs) are characterized by a two-fold process: a forward process and a backward process. The forward process operates by incrementally introducing Gaussian noise into the data according to schedule \(\beta_{1}, \ldots, \beta_{T}\). 
\begin{equation}
q\left(x_{1: T} \mid x_{0}\right):=\prod_{t=1}^{T} q\left(x_{t} \mid x_{t-1}\right) \\
\end{equation}
\begin{equation}
q\left(x_{t} \mid x_{t-1}\right):=\mathcal{N}\left(x_{t} ; \sqrt{1-\beta_{t}} x_{t-1}, \beta_{t} \mathbf{I}\right)
\end{equation}

The forward process facilitates the sampling of the data  \(x_{t}\) at an arbitrary timestep \(t\) based on the closed form as:
\begin{equation}
q\left(x_{t} \mid x_{0}\right):=\mathcal{N}\left(x_{t} ; \sqrt{\bar{\alpha}_{t}} x_{0},\left(1-\bar{\alpha}_{t}\right) \mathbf{I}\right)
\end{equation}
Equivalently:
\begin{equation}
x_{t}:=\sqrt{\bar{\alpha}_{t}} x_{0}+\sqrt{1-\bar{\alpha}_{t}} \epsilon, \quad \text { where } \epsilon \sim \mathcal{N}(\mathbf{0}, \mathbf{I})
\end{equation}
where \(\alpha_{t}:=1-\beta_{t}\) and \(\bar{\alpha}_{t}:=\) \(\prod_{s=1}^{t} \alpha_{s}\). 

The backward process is essential for iteratively recovering information, thereby enabling the generation of new data from random Gaussian noise. The key parameter in this process is \(\beta_{t}\), representing the noise variance at each timestep. When \(\beta_{t}\) is small, the reverse step aligns with a Gaussian distribution, facilitating the gradual denoising of the data.

\begin{equation}
p_{\theta}\left(x_{0: T}\right):=p\left(x_{T}\right) \prod_{t=1}^{T} p_{\theta}\left(x_{t-1} \mid x_{t}\right)
\end{equation}
\begin{equation}
p_{\theta}\left(x_{t-1} \mid x_{t}\right):=\mathcal{N}\left(x_{t-1} ; \tilde{\mu}_{t}, \tilde{\beta}_{t} \mathbf{I}\right)
\end{equation}
where variance \(\tilde{\beta}_{t}\) can be calculated from the forward process posteriors:
$\tilde{\beta}_{t}:=\frac{1-\bar{\alpha}_{t-1}}{1-\bar{\alpha}_{t}} \beta_{t}$

Following method proposed in \cite{lin2024common} which has shown improvements in audio generation \cite{hai2023dpm}, we apply a fixed noisy schedule to $\beta_{t}$ and $\alpha_{t}$ and use velocity $v_t$ instead of noise $\epsilon$ as the neural network's prediction target:

\begin{equation}
v_{t}:=\sqrt{\bar{\alpha}_{t}} \epsilon-\sqrt{1-\bar{\alpha}_{t}} x_{0}
\end{equation}

According to (4) and (7), the backward process is then performed by the following functions:
\begin{equation}
x_{0}:=\sqrt{\bar{\alpha}_{t}} x_{t}-\sqrt{1-\bar{\alpha}_{t}} v_{t}
\end{equation}
\begin{equation}
\tilde{\mu}_{t}:=\frac{\sqrt{\bar{\alpha}_{t-1}} \beta_{t}}{1-\bar{\alpha}_{t}} x_{0}+\frac{\sqrt{\alpha_{t}}\left(1-\bar{\alpha}_{t-1}\right)}{1-\bar{\alpha}_{t}} x_{t}
\end{equation}

CFG \cite{ho2022classifier} is increasingly being adopted to steer the sampling process in diffusion models. This technique modifies the model output $v$ during sampling, as described by the equation:

\begin{equation}
v_{cfg} = v_{neg} + w (v_{pos}-v_{neg}) 
\end{equation}
 where $w$ represents the guidance scale, and $v_{pos}$ and $v_{neg}$ denote the model outputs under positive and negative conditions, respectively. And $v_{cfg}$ is the classifier-free guided velocity.
 
 To further refine this process, a rescaling method proposed in \cite{lin2024common} is applied to $v_{cfg}$ to enhance its effectiveness and mitigate over-exposure when $w$ is large.

\begin{equation}
v_{re} = v_{cfg} \cdot \frac{std(v_{pos})}{std(v_{cfg})} \\
\end{equation}

\begin{equation}
v'_{cfg}=\phi \cdot v_{re}+(1-\phi) \cdot v_{c f g}
\end{equation}

Here, $\phi$ is a hyperparameter used to control the strength of the rescale adjustment. $v'_{cfg}$ is the rescaled CFG velocity used for diffusion sampling. 

\begin{table*}[h]
 \small
  \caption{Comparison of Objective scores: Word Error Rate (WER), Phoneme Error Rate (PER), Relative Inference Speed (RIS), and Mean Opinion Scores (MOS) with their 95\% confidence intervals (CI): Q-Quality, N-Naturalness, C-Prompt-Voice-Consistency.}
  \label{tab:result1}
  \footnotesize
  \centering
  \setlength{\tabcolsep}{10pt} 
  \begin{tabular}{c|c|ccc|ccc}
    \hline
    \textbf{Method}&\textbf{Text-Guided VC}&\textbf{WER $\downarrow$}&\textbf{PER $\downarrow$}&\textbf{RIS$\uparrow$}&\textbf{MOS-Q $\uparrow$}&\textbf{MOS-N $\uparrow$}&\textbf{MOS-C $\uparrow$}\\
    \hline
    Grount-Truth & / & / & / & / & $4.42\pm0.11$ & $4.26\pm0.11$ & $4.12\pm0.13$  \\
    \hline
    FreeVC & $\times$ & $6.37$ & $9.79$ & / & $4.09 \pm 0.12$ & $3.98 \pm 0.13$ & / \\
    ReDiffVC & $\times$ & $3.45$ & $8.26$ &  /&$3.67 \pm 0.14$ & $3.76 \pm 0.13$ & / \\
    \hline
    DreamVC & $\checkmark $ & \textbf{4.10} & \textbf{8.08} & 1.00x & $3.62\pm0.14$ & $3.61\pm0.14$ & \textbf{3.72 $\pm$ 0.15} \\
    DreamVG+FreeVC & \checkmark & $7.58$ & $10.05$ & 2.71x & \textbf{3.90 $\pm$ 0.13} & \textbf{3.85 $\pm$ 0.14} & $3.43\pm0.16$ \\   
    DreamVG+ReDiffVC & \checkmark & $5.11$ & $8.65$ & 1.08x &  $3.80\pm0.14$ & $3.70\pm0.13$ & $3.66\pm0.15$ \\
    \hline
  \end{tabular}
  \label{table1}
\end{table*}


\subsection{DreamVC: Text-to-Voice Conversion Model}


The DreamVC model leverages a text-guided process to modify the timbre of the source speech based on the given text prompt. The model is based on a conditional diffusion model that uses speech content 
and text prompt 
as dual conditions to guide the generation of the output as detailed in Figure \ref{fig:model}(a). The output mel-spectrogram is converted to waveform using the pre-trained neural Vocoder. This model distinguishes itself from DiffVC \cite{popov2021diffusion} in several key aspects: it eschews the use of an average-voice encoder and instead uses a pre-trained SLM for disentangling voice and content, integrates cross-attention layers to merge text prompts effectively, and employs the CFG to control the impact of conditions.

\subsection{DreamVG: Text-to-Voice Generation Plugin}

However, as an end-to-end text-guided voice conversion model based on the diffusion model, DreamVC faces limitations in real-world applications due to its drawbacks such as slow inference speed, high memory usage, expensive training, and difficulty in reproducing a desirable voice that was once generated. To address these issues, we introduce DreamVG, an alternative model that adopts a plug-and-use strategy. DreamVG efficiently generates latent speaker embeddings from text prompts using a conditional diffusion model, enhancing its practicality and application scope, as illustrated in Figure \ref{fig:model}(b). This module can act as a replacement for any one-shot VC model that uses latent speaker embeddings to generate the target voice, as shown in Figure \ref{fig:model}(c). The plugin method of DreamVG boosts the functionality of pre-trained one-shot voice conversion models, rendering it a flexible solution to enable text guidance.





\section{Experiments}



\subsection{Experimental Settings}
For this study, we used the recordings of 900 speakers from VCTK \cite{Yamagishi2019VCTK} and LibriTTS-R \cite{koizumi2023libritts} datasets and their text prompts from the proposed annotated dataset DreamVoiceDB as mentioned in Section \ref{sec:dreamvoicedb} for training, and used speakers from LibriTTS-R dev set for validation and test.
All the audio files were sampled at 24KHz for synthesis and 16KHz for content and speaker embedding extraction.

The U-Net model \cite{ronneberger2015u} used in DreamVC has 3 downsampling and 3 upsampling blocks configured with 128, 256, and 512 channels respectively, and each of the 4 blocks in the middle has a cross-attention layer, totaling 103.2M parameters. The pre-trained T5 base \cite{raffel2020T5}, noted for its excellence in various generative tasks, is used to process text prompts. The pre-trained ContentVec \cite{qian2022contentvec} is used for content embedding extraction and a pre-trained BigVGAN \cite{lee2022bigvgan} is employed as the neural vocoder. Based on the configuration of the BigVGAN vocoder, the mel-spectrogram has 100 mel-spectrograms, a window size of 1024, and a hop size of 256.
Embeddings extracted from ContenVec are duplicated by hard mapping to match the sample rate of mel-spectrogram. 
The diffusion steps and inference steps for the default DreamVC are 1000 and 50 respectively, and the corresponding variance $\beta$ is set from 0.0001 to 0.02. During sampling, we found the value of guidance scale $w$ as 3, a rescaling factor $\phi$ of 0.7, and setting an unconditional prompt as the negative condition can lead to better generation quality.

The neural network in DreamVG has three blocks configured with 128, 256, and 256 channels, where each block has a cross-attention layer, totaling 26.2M parameters. Similar to DreamVC, we use the T5 base model to generate the prompt embeddings. We adopted the speaker verification model commonly applied in one-shot VC models \cite{li2023freevc, liu2021any} as the model output for DreamVG.
The diffusion steps and inference steps for the default DreamVG are 1000 and 100 respectively, and the corresponding variance $\beta$ is set from 0.0001 to 0.02. During sampling, we use a guidance scale $w$ of 3 and a rescaling factor $\phi$ of 0.7.
The DreamVG is integrated with two pre-trained one-shot VC models to facilitate text-guided control in VC: (1) FreeVC \cite{li2023freevc} is one of the state-of-the-art one-shot voice conversion models that utilizes the VITS \cite{kim2021conditional} architecture enhanced by GAN training. (2) ReDiffVC is a variation of DreamVC designed for one-shot voice conversion, which replaces cross-attention blocks with self-attention blocks. ReDiffVC incorporates the one-shot speaker embedding by adding it to the diffusion step embedding. Additionally, it employs CFG with a guidance scale of 3 and a rescaling factor of 0.7, using an empty speaker embedding as the negative condition during sampling. 

\subsection{Evaluation Metrics}

 For objective evaluation, we chose 120 utterances from the LibriTTS-R development set, each with a new, randomly generated text prompt. For subjective evaluations, we selected 15 utterances from this set, each with a unique, manually created sentence prompt. In addition, 15 targets from the training set with their prompts are also used for MOS evaluation.

The proposed models were evaluated to assess their quality (MOS-Q), naturalness (MOS-N), prompt-voice consistency (MOS-C), and relative inference speed (RIS). Specifically, for MOS-C, listeners evaluated how well the generated speaker's timbre in synthetic speech aligned with provided text descriptions, assigning scores ranging from 1 (total mismatch) to 5 (perfect match). The MOS-Q assesses the quality of synthesized audio, focusing on aspects like noise and artifacts, while the MOS-N evaluates the naturalness of the generated voice.
A total of 15 English speakers, hired online, participated in these subjective evaluations. In addition to the above subjective measures, objective intelligibility of the converted speech is estimated using WER and PER from the words and phonemes transcribed by Whisper-medium \cite{radford2023robust} and Allosaurus \cite{li2020universal} respectively.

\subsection{Experimental Results}

As shown in the Table \ref{table1},
the DreamVG+FreeVC demonstrated a higher WER and PER compared to DreamVC, indicating a decrease in the intelligibility. Conversely, for subjective aspects like voice quality and naturalness, the combination of DreamVG and FreeVC outperformed others. This is likely due to FreeVC's simpler single-stage process, as opposed to the more complex two-stage Diffusion-based model that relies heavily on a Neural Vocoder trained independently. The gap in naturalness and sound quality of ReDiffVC and DreamVC could also be caused by the mismatch issue of content embedding's sample rate and mel-spectorgram's sample rate.

DreamVC is superior in maintaining prompt-voice consistency in comparison with DreamVG plugin based methods. DreamVG+FreeVC exhibited the lowest MOS-C, likely because FreeVC struggled to effectively disentangle speaker information from the content embedding. The superior performance of DreamVC can be attributed to the effectiveness of CFG in removing speaker information from source more effectively. 

Interestingly, combining the plugin method with a very light VC variant resulted in faster inference speeds when compared to the more comprehensive end-to-end DreamVC approach. However, this combination might slightly compromise the performance of the one-shot voice conversion model. 

\section{Conclusion and Future Work}


In conclusion, our research introduces the DreamVoiceDB, a comprehensive, open-source voice timbre dataset annotated by professionals that can be leveraged as a valuable resource for the next generation of generative voice applications. In addition, we propose two novel text-guided voice generation models utilizing DPM and CFG: DreamVC, an innovative voice timbre conversion model for enhanced controllability, and DreamVG, a versatile text-to-voice generation plugin model, compatible with one-shot VC models for speaker embeddings generation. 

Our experiments demonstrate the effectiveness of these models and the proposed dataset in generating voice timbres precisely aligned with text descriptions. While DreamVG shows adaptability in integration with other models, DreamVC performs decently on prompt-voice consistency and suffers from sound quality and naturalness issues. This can be attributed to the sampling rate mismatch between the neural vocoder and content encoder, and the complexities of its two-stage structure without directly optimizing the waveform signal. In the future, there is considerable scope for improving the overall performance and the speed of these models to enable the generation of one's \textit{``DreamVoice''}.

\newpage

\bibliographystyle{IEEEtran}
\bibliography{mybib}

\end{document}